\documentclass[amsmath,amssymb,reprint,twocolumn, prb,superscriptaddress ]{revtex4-1}

\usepackage[utf8]{inputenc}
\usepackage[T1]{fontenc}

\usepackage[english]{babel}

\usepackage{amsmath,amssymb,amsfonts}
\usepackage{amstext, mathrsfs, textcomp}
\usepackage{mathptmx}
\usepackage{bigints}  
\usepackage{nicefrac}
\usepackage{multirow}
\usepackage{dcolumn}
\usepackage{bm}
\usepackage[separate-uncertainty = true,multi-part-units=single]{siunitx}	

\usepackage{booktabs}
\usepackage{subfigure}
\usepackage{graphicx}
\usepackage{xcolor}
\usepackage[export]{adjustbox}

\usepackage[colorlinks=true, citecolor=blue, urlcolor=blue, linkcolor=blue]{hyperref}

\usepackage{changes}

\graphicspath{{Figures/}{supinfo/Figures/}}

\newcommand{\TITLE}{Deep learning enabled design of complex transmission matrices for universal optical components}

\newcommand{\missingref}[1]{\textcolor{red}{(ADD REF!)}}

\begin{document}

\title{\TITLE}

\author{\firstname{Nicholas J.} \surname{Dinsdale}}
\affiliation{Optoelectronics Research Centre, University of Southampton, Southampton, UK}
\affiliation{Physics and Astronomy, Faculty of Engineering and Physical Sciences, University of Southampton, SO17 1BJ Southampton, UK}

\author{\firstname{Peter R.} \surname{Wiecha}}
\email[e-mail~: ]{pwiecha@laas.fr}
\affiliation{Physics and Astronomy, Faculty of Engineering and Physical Sciences, University of Southampton, SO17 1BJ Southampton, UK}
\affiliation{LAAS, Universit\'e de Toulouse, CNRS, Toulouse, France}

\author{\firstname{Matthew} \surname{Delaney}}
\affiliation{Optoelectronics Research Centre, University of Southampton, Southampton, UK}
\affiliation{Physics and Astronomy, Faculty of Engineering and Physical Sciences, University of Southampton, SO17 1BJ Southampton, UK}

\author{\firstname{Jamie} \surname{Reynolds}}
\affiliation{Optoelectronics Research Centre, University of Southampton, Southampton, UK}

\author{\firstname{Martin} \surname{Ebert}}
\affiliation{Optoelectronics Research Centre, University of Southampton, Southampton, UK}

\author{\firstname{Ioannis} \surname{Zeimpekis}}
\affiliation{Optoelectronics Research Centre, University of Southampton, Southampton, UK}

\author{\firstname{David J.} \surname{Thomson}}
\affiliation{Optoelectronics Research Centre, University of Southampton, Southampton, UK}

\author{\firstname{Graham T.} \surname{Reed}}
\affiliation{Optoelectronics Research Centre, University of Southampton, Southampton, UK}

\author{\firstname{Philippe} \surname{Lalanne}}
\affiliation{LP2N, CNRS - Institut d’Optique Graduate School - Univ. Bordeaux, F-33400 Talence, France}

\author{\firstname{Kevin} \surname{Vynck}}
\affiliation{LP2N, CNRS - Institut d’Optique Graduate School - Univ. Bordeaux, F-33400 Talence, France}

\author{\firstname{Otto L.} \surname{Muskens}}
\email[e-mail~: ]{o.muskens@soton.ac.uk}
\affiliation{Physics and Astronomy, Faculty of Engineering and Physical Sciences, University of Southampton, SO17 1BJ Southampton, UK}

\begin{abstract}
Recent breakthroughs in photonics-based quantum, neuromorphic and analogue processing have pointed out the need for new schemes for fully programmable nanophotonic devices. Universal optical elements based on interferometer meshes are underpinning many of these new technologies, however this is achieved at the cost of an overall footprint that is very large compared to the limited chip real estate, restricting the scalability of this approach. Here, we consider an ultracompact platform for low-loss programmable elements using the complex transmission matrix of a multi-port multimode waveguide. We propose a deep learning inverse network approach to design arbitrary transmission matrices using patterns of weakly scattering perturbations. The demonstrated technique allows control over both the intensity and phase in a multiport device at a four orders reduced device footprint compared to conventional technologies, thus opening the door for large-scale integrated universal networks.
{\newline\textbf{keywords:} silicon photonics, deep learning, inverse design, light scattering}
\end{abstract}

\maketitle

\section*{Introduction}

Programmable photonic integrated circuits have been introduced as a new paradigm for the datacommunications domain, to circumvent the limitations of application-specific fabrication that is required for even the smallest changes in design \cite{Bogaerts2020}. Reconfigurable components are finding immediate applications in post-fabrication trimming  \cite{Chen2017}, tunable elements for microwave photonics \cite{Zhuang2015}, and optical switch matrices \cite{Stabile2016}. However, emerging areas are on the horizon and quantum and neuromorphic computing have been recently seeing breakthrough developments \cite{Shen2017,Harris2017,Harris2018, Wang2020}. The conventional engineering toolkit for such reconfigurable photonic networks consists of interferometer meshes and directional couplers, which have been successfully applied to achieve universal optical elements \cite{Miller2015, Clements2016}. Ultimately, an approach based on cascading of many single-mode devices to obtain a multi-mode logical circuit will face the limited real estate available on a silicon chip, limiting the scalability of this approach.

Generally, the input-output relationship of linear optical elements is fully described by its transmission matrix, which is a subpart of the system's scattering matrix,\cite{popoffMeasuringTransmissionMatrix2010} and multiple scattering allows strong mixing of input and output ports \cite{Rotter2017}. Here, each output port contains the full information of all inputs in a scrambled manner, with for example applications in compressive sensing. Control over the transmission can be achieved via the shaping of the phase and amplitudes of the input degrees of freedom, thereby allowing the descrambling of the information from a multiple scattering pattern \cite{Mosk2012}. Alternatively one can control the transmission matrix of the medium itself by carefully positioning the individual scatterers, however, this is a problem which quickly diverges with the number of scattering sites. Nevertheless, scattering devices that are based on complex nanostructuring of a photonic waveguide have been proposed that can be optimized for a particular task \cite{Liu2011, Jensen2011, luNanophotonicComputationalDesign2013,piggott2015inverse, frellsenTopologyOptimizedMode2016}. In these devices, strong multiple scattering of photons is used to achieve ultracompact device footprints. Similar complex structured devices are currently being considered for emerging applications in analogue computing \cite{Hughes2019,estakhriInversedesignedMetastructuresThat2019}. The basic principle here is that the complex medium allows the modeling of a specific mathematical function that performs a linear operation on an input signal.

As an intermediate solution between strong multiple scattering designs and weaker diffractive optical elements, the application of patterns of small perturbations to a multimode interference device was recently shown to result in a programmable integrated spatial light modulator on silicon\cite{bruckAllopticalSpatialLight2016}. The weak perturbation limit is very different from the strong scattering-based complex devices and is more related to, for example, branched-flow scattering systems \cite{brandstotterShapingBranchedFlow2019}. A particular advantage is that weak perturbations allow controlling the flow of light with very high overall throughput and minimal reflection losses.

Here, we demonstrate that weakly perturbed multimode waveguides in fact allow full control over the transmission matrix, enabling the design of universal optical components. We successfully address the challenge of inversely designing the scattering medium by using a neural network approach. Deep learning\cite{goodfellowDeepLearning2016} has proven in the last few years to offer powerful tools for data processing and solving of problems with tremendous complexity, and has ever since been considered of highest relevance for photonics by the scientific community. Various impressive applications have been reported in (nano-) photonics like phase recovery in optical microscopy,\cite{rivensonPhaseRecoveryHolographic2018} optical characterization and classification of nanometer-scale specimen \cite{joHolographicDeepLearning2017, wiechaPushingLimitsOptical2019}, interpretation and discovery of implicit physical mechanisms\cite{yeungElucidatingBehaviorNanophotonic2020, kiarashinejadDeepLearningReveals2019} or the real-time interpretation of light scattering through complex media \cite{borhaniLearningSeeMultimode2018}.

Particular research interest has been put into what is usually called \textit{inverse design}, the design of a device offering a specific, pre-defined functionality. There have been numerous demonstrations of etched silicon perturbation patterns for wavelength, modal and polarization multiplexing \cite{piggott2015inverse,frellsenTopologyOptimizedMode2016,shenIntegratednanophotonicsPolarizationBeamsplitter2015}. Unfortunately, even for the most simple case of a binary perturbation pattern the problem contains $2^P$ possible solutions, where $P$ is the number of possible perturbation positions. The search for appropriate patterns is therefore usually based on topological optimization, evolutionary algorithms or other computationally very expensive heuristics \cite{feichtnerEvolutionaryOptimizationOptical2012, elsawyNumericalOptimizationMethods2020,bruckAllopticalSpatialLight2016, xu2017integrated, jia2018inverse, van2019approximating, xieDesignArbitraryRatio2020}.

A deep artificial neural network (ANN) can in principle be trained on the solution of inverse problems and can, in particular, learn to inverse design nano-photonic devices\cite{hegdeDeepLearningNew2020, tahersimaDeepNeuralNetwork2019, jiangDeepNeuralNetworks2020, zhelyeznyakovDeepLearningAccelerate2020, kiarashinejadDeepLearningApproach2019}.
Transmission matrix design using sub-wavelength refractive index perturbations represents an ideal challenge for this novel ANN approach, owing to the complexity of the patterns required and the lack of a fast numerical solution for pattern optimization. In this work we leverage the power of full-wave Maxwell's equations solvers to generate a library of solutions for random perturbation patterns within multimode interference devices (MMIs) with different port configurations, and train a neural network to be able to predict perturbation patterns, in real time, to create an arbitrary transmission matrix.
We demonstrate that our ANN approach allows not only to tailor incoherent transmitted intensities, but also can be used to gain independent control over the phase in each transmission channel. The full control over matrix elements enables rational design of universal optical elements and multi-port unitary operators.

\section*{Results}

\begin{figure*}[t]
	\centering
	\includegraphics[width=0.95\linewidth]{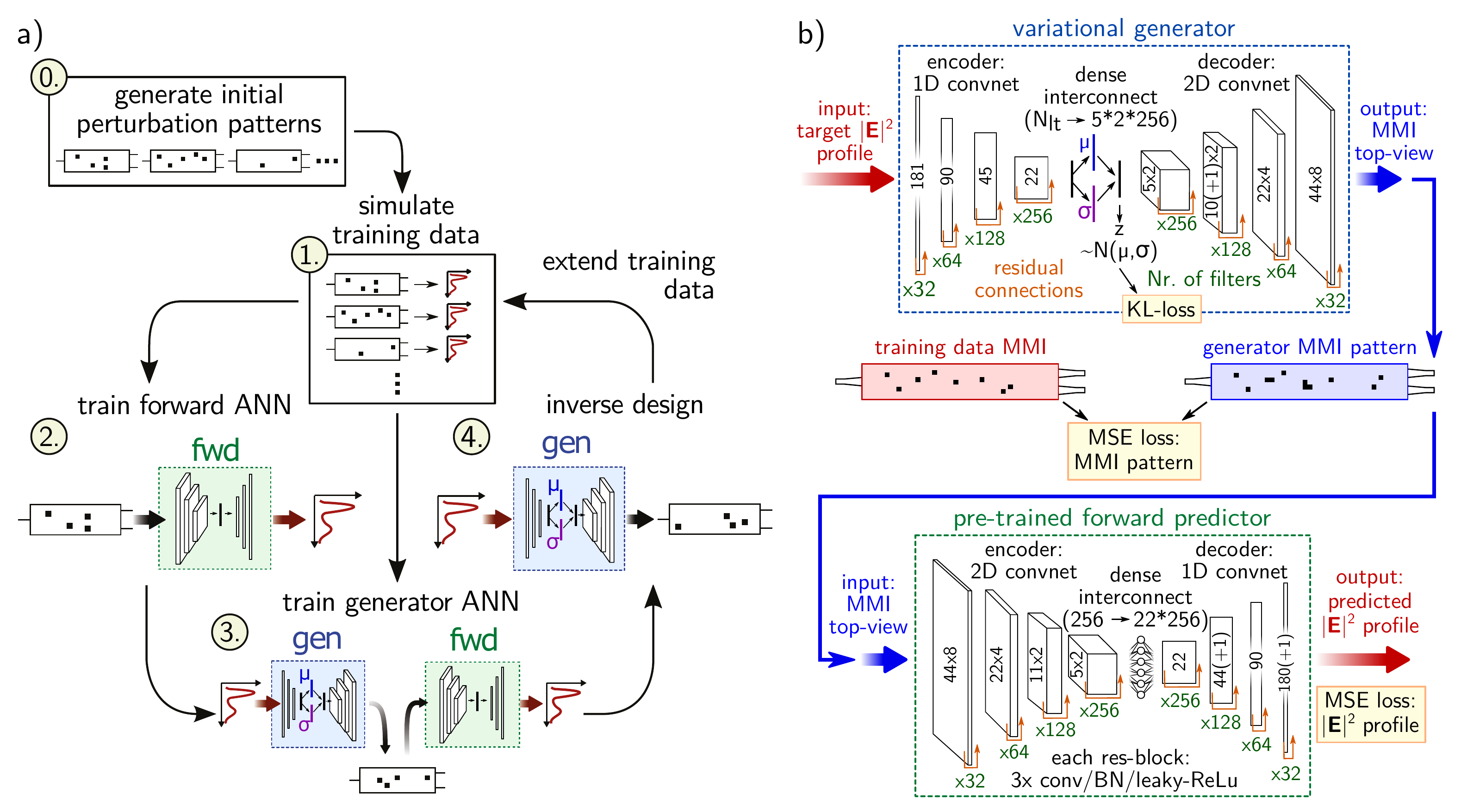}
	\caption{(a) scheme of the iterative data generation procedure.
		An initial dataset is generated, simulated and subsequently used for ANN-training.
		First the forward network is trained using only the intensity-profile loss. The forward network's weights are then fixed and the generator network is trained.
		The trained generator ANN is then used to inverse-design new patterns, which are simulated and added to the training data.
		(b) detailed overview of the full inverse-design ANN model, consisting of variational generator (blue box) and pre-trained forward predictor network (green box).
		The full model is trained using the sum of three losses: MMI-pattern loss, $|\mathbf{E}|^2$-loss and KL-loss with weightings $1 : 50 : 0.2$, respectively.
	}
	\label{fig:training_scheme}
\end{figure*}

\subsection*{Problem, Model and Neural Network Architecture}

\paragraph*{MMI device model.} 
As a demonstration of capabilities, we consider $M$-input by $N$-output MMI rib waveguides with lateral dimensions of $33\times 6\,$\textmu m$^2$ obtained by etching 120\,nm into a $220$\,nm thick silicon layer of an silicon-on-insulator (SOI) wafer. Details are presented in the Methods section. The device is patterned by non-overlapping square perturbations of $750\times 750\,$nm$^2$ in size, resulting in a 8$\times$44 grid of possible positions.
Consequently, our geometry allows $2^{352}$ (around $10^{106}$) possible different patterns, a number which renders a systematic search for specific MMI designs impossible.
The input and output waveguides are single mode with a width of $500$\,nm and taper up to $1$\,\textmu m wide at the MMI region boundary over a length of $10$\,\textmu m, thereby allowing the adiabatic expansion of the fundamental TE mode and reducing transmission losses~\cite{thomson2010low}. In our studies we focus on two cases, respectively corresponding to $1\times 2$ and $3 \times 3$ multi-port MMIs. In each case, the waveguides are centered with respect to the MMI and are separated by $3$\,\textmu m (in case of 2 ports), respectively $2.15$\,\textmu m (in case of 3 ports).
In our experiment both the rib waveguides and the perturbation patterns are defined using a single electron beam lithography step and are therefore self-aligned. We note however that our wave shaping approach is fairly robust to fabrication tolerances (as shown in SI Fig.~S11) and the structure sizes are sufficiently large so that it can be implemented using standard $193$\,nm deep UV lithography, without having to rely on ultra precision nanofabrication, as is the case for many other inverse designs.

\paragraph*{Training data generation for arbitrary transmission states.} 
A systematic evaluation of all $2^{352}$ possible patterns is impossible. However, only a very sparse subset of these patterns provides a high overall throughput as light is usually scattered out of the device and does not arrive at the output waveguides. Consequently, a network trained on exclusively random perturbation patterns will not manage to generalize to those desired patterns which result in high throughput solutions with specific splitting ratios. Therefore, in order to improve the quality of the solutions, it is critical that the network is trained on a set of solutions with reasonably high throughput.
This is why prior to the training of the actual ANN we first need to develop a method to generate a training set of high-performance MMIs, which still contain a sufficient amount of randomness in their perturbation layouts.

The overall scheme of our data generation process is presented in Fig.~\ref{fig:training_scheme}a.
Randomized splitting ratios are chosen as target values, using a normalization scheme for the total transmittance (see supporting information). The MMI pattern is then optimized towards the target through an iterative process by maximizing a fitness function as described in detail in the Supporting Information. In short, in this process perturbations are added successively one-by-one in a randomized order of positions for the grid. Each pattern is evaluated by a full-field numerical Maxwell solver. If the fitness function is improved once a new perturbation was added, this perturbation is kept. Perturbations that do not improve the throughput are discarded. The process is repeated until one of following stop criteria is met: either the fitness function $f$ cannot be further improved, or the maximum allowed number of $50$ accepted perturbations is reached. Practically, on average the fitness cannot be further improved after around $30$ accepted perturbations. Figure~S1 in the supporting information illustrates an example of pattern generation for the optimization of coupling towards a single output of a $1\times 2$ MMI.

For the initial pattern generation (step 0), we use an in-house implementation of the aperiodic Fourier Modal Method (a-FMM)~\cite{silberstein2001use, vynckUltrafastPerturbationMaps2018}. The a-FMM relies on a supercell method and perfectly matched layers to compute the modes of the system, and a scattering matrix formalism to describe mode coupling and has previously been shown to provide numerical predictions with high accuracy and fast convergence~\cite{lalanne2007numerical}. It has also been demonstrated to be well-suited to the study of multi-port MMI devices containing perturbations~\cite{bruckAllopticalSpatialLight2016,vynckUltrafastPerturbationMaps2018}, owing to its very fast runtime and direct access to the full set of coupling coefficients between all input and output waveguide modes. Simulations are performed in 2D using the effective index method~\cite{chrostowski2015silicon}. 
The accuracy of the 2D simulations is illustrated in the supporting information section~H by a comparison with 3D simulations (Fig.~S10). In the supporting information section~I (tables~S2 and~S3) we provide a discussion on expected scattering losses.

Subsequently, in step 1 of Fig.~\ref{fig:training_scheme}a we simulate all accepted patterns using a finite-difference time-domain (FDTD) approach (Lumerical FDTD), using the same parameters as with the a-FMM approach. This numerical method allows closer comparison to the experimental situation, as the tapered input and output waveguides cannot be efficiently modelled with the a-FMM. The electric field intensity profiles across the single-mode output waveguides are calculated at a distance of 20\,\textmu m from the MMI end plane, sufficiently far away from the taper. For each pattern, the transmittance profiles are simulated with an incoming fundamental TE mode into each input port at $\lambda_0=1550$\,nm. Finally, the size of the dataset is doubled by exploiting the symmetry of our systems about the propagation axis.

The initial datasets comprise of a total of 2500 of MMI-``families'', where each family corresponds to the ensemble of accepted iterations in an optimization cycle. After exploiting the mirror symmetry, the generated datasets consist of around $1.5 \times 10^5$ patterns.

\paragraph*{Tandem neural network layout.} 
We use a network model based on a combination of two encoder-decoder type ANNs. Each network is composed of so-called ``residual'' down- and up-sampling convolutional blocks.
Such ``ResNets'' allow to maintain efficient training convergence of very deep network layouts \cite{szegedyInceptionv4InceptionResNetImpact2016}.

The first part for our ANN model is a forward network, which acts as a ``physics predictor''. A detailed overview of the predictor network structure is given by the green box in figure~\ref{fig:training_scheme}b.
It takes as input a 2D perturbation pattern and returns the predicted 1D output intensity profile.
We use 90\% of the dataset for the actual training and keep 10\% for validation. Trained on this initial training data, the forward network achieves reconstruction of the field intensity at the output ports with a $<$5\% average error.
Once trained, the forward-ANN parameters are fixed during the subsequent inverse-network training step.

In step 3 of figure~\ref{fig:training_scheme}a, we introduce the second component of our model, the generator ANN. Details are given by the blue box in Fig.~\ref{fig:training_scheme}b. The generator takes the inverse design target as input, which is here the intensity profile across the output waveguides of the MMI. In the case of multiple input channels, the ANN takes a separate intensity target for each input. The generator ANN returns an MMI perturbation pattern.

\begin{figure*}[tb]
	\centering
	\includegraphics[width=.9\linewidth]{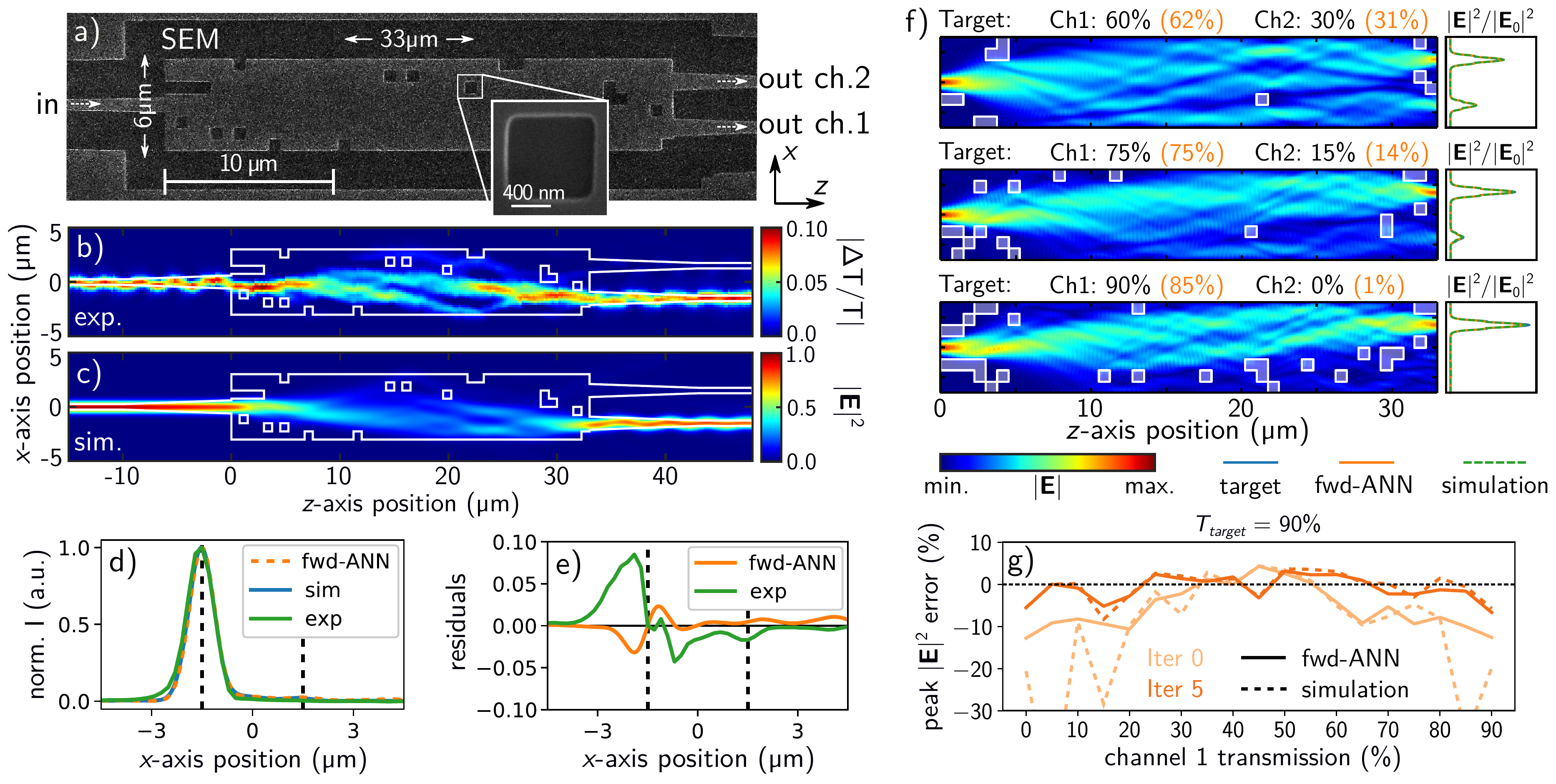}
	\caption{
		(a-d) Manually designed $1\times 2$ etched MMI structure for comparison of experimental conditions with the 2D simulations and with the forward ANN predictions.
		(a) MMI geometry, scale bars represent $10\,$\textmu m (main) and $400\,$nm (zoom).
		(b) perturbation-based transmittance measurement. \cite{bruckDevicelevelCharacterizationFlow2015, vynckUltrafastPerturbationMaps2018} (c) field intensity distribution FDTD simulation.
		(d) illustration of the forward ANN which we test against the simulations and the experiment.
		(e) intensity-profile at output, obtained experimentally by perturbation-measurement (green), by FDTD simulation (blue) and with the forward neural network (orange).
		For the comparison, simulated and forward-ANN predicted intensity profiles in (c) and (d) are convoluted with a Gaussian of width 740\,nm corresponding to the laser spot-size.
		(f) examples of inverse designed etched $1\times 2$ MMI patterns (white overlay) with target 90\% total transmittance and different splitting ratios. The electric field distributions of the etched patterns are calculated by FDTD simulations. Target and simulated transmittance values are given by the black and orange numbers, respectively. Right side of (f) shows the intensity profiles along the output waveguides for design target (solid blue lines), forward-ANN prediction (solid orange lines) and FDTD simulation (dashed green lines).
		(g) Inverse design MMI peak intensity error for forward-ANN prediction (solid line) and simulation (dashed line) verses target with respect to different channel splitting ratios and a total device transmittance of 90\%. Pale and dark colors correspond to initial and final network iterations, respectively.
	}
	\label{fig:experiment_1x2}
\end{figure*}

For the training, predicted  MMI patterns are compared with the training patterns (``MMI pattern'' loss in figure~\ref{fig:training_scheme}b). However, it has been shown that this direct training is not sufficient to solve ``many-to-one'' problems such as the present one, where different patterns may lead to an identical (or very similar) transmission through the MMI and hence the solution is non-unique.\cite{liuTrainingDeepNeural2018}.
To avoid that the network gets confused by different solutions to the same problem, the generated MMI patterns are fed back into the pre-trained forward ANN. Its predicted transmittance is compared to the training data, yielding an ``$|\mathbf{E}|^2$-profile'' loss, associated with the predicted MMI perturbation pattern.
Such tandem network architecture has recently proven to be very successful in photonics inverse design \cite{liuTrainingDeepNeural2018, xuEnhancedLightMatterInteractions2019, soSimultaneousInverseDesign2019, maProbabilisticRepresentationInverse2019}. Details on our ANN model can be found in the Methods section.

Our generator network is implemented as an encoder-decoder adaptation of a $\beta$-variational autoencoder ($\beta$-VAE) \cite{i.higginsVVAELearningBasic2017, burgessUnderstandingDisentanglingBeta2018, kingmaIntroductionVariationalAutoencoders2019}.
This is done by adding a Kullback-Leibler-divergence term as a third loss to the training of the full model (``KL-loss'' in figure~\ref{fig:training_scheme}b). 
While more details are given in the Methods section, the KL-loss basically forces the network to learn a smooth and continuous compressed representation (center ``dense interconnect'' layer in the network schemes in figure~\ref{fig:training_scheme}b). It prevents that a small change in the design target unexpectedly leads to a radically different MMI pattern and hence helps stabilizing the inverse design robustness.
More details are given in the Methods section, typical plots of training convergence are given in the supporting information figure~S2.

Since the training of the ANNs relies on calculating gradients of the network output with respect to its parameters, the generated MMI patterns are necessarily gray-scale and need to be converted to a binary design. We do this by applying a threshold below which all gray-scale values are set to zero and the remaining positions are set to one. In order to avoid choosing an arbitrary threshold value, we use the pre-trained forward network to evaluate the resulting MMIs for a series of threshold values between 0 and 1 (in steps of 0.01). Eventually we choose the solution closest to the design target, using a mean square error metric. For more details and an analysis of the impact of varying the threshold, see supporting information section~1.2 and SI figure~S3. Thanks to the very short prediction time of less than 0.5\,ms per MMI (on a NVIDIA Quadro P6000 GPU), this does not create a considerable computational bottleneck. We still obtain the final inverse design in a time of the order of 50 milliseconds. We note that inference is also relatively efficient on CPU, the full process taking around 130\,ms on a 3rd generation AMD Ryzen processor.

\paragraph*{Iterative training.} 
From initial investigations of inverse design performance (see supporting information figure~S4 and ``iter 0'' in figure~\ref{fig:experiment_1x2}g which is discussed further below), we hypothesize that the homogeneity of the training data, which comprises MMIs that are produced by the same data-generation algorithm, hinders the forward network's ability to correctly generalize. Since the forward ANN plays a crucial role in the training of the generator ANN itself, the accuracy of the forward ANN poses a fundamental problem for the inverse design. We therefore decided to implement an iterative training scheme based on the extension of the training dataset with ANN-generated patterns. This helps to improve the diversity of the training data and incorporates a self-correcting feedback effect into our network training scheme, thereby teaching the network in a ``self-consistent'' way to avoid its previous errors. Similar training techniques have been recently proposed for related inverse design problems \cite{wenProgressiveGrowingGenerativeAdversarial2019, blanchard-dionneSuccessiveTrainingGenerative2020a}.

In our iterative scheme depicted in Figure~\ref{fig:training_scheme}a, after each training cycle (steps 1-3), we use the generator network in step 4 to produce a dataset of 10000 random, inverse designed perturbation patterns. For details on the randomized inverse target generation, see the supporting information section~1.1. We simulate the new inverse designed patterns by FDTD and combine these results with the previous training set, effectively closing the loop between step 4 and step 1.
On the combined data, we train again first the forward ANN and then the generator network in the same way as described above. Through the ANN-generated, FDTD-simulated data, any features that were previously poorly modeled are fed back into the ANN model, thereby improving its performance.
For all models in this work, we repeated the iterative data generation process over 6 full cycles. Detailed statistics of the networks during this iterative process can be found in the supporting information figures~S4 and~S5 for the $1\times 2$ and $3\times 3$ MMI cases, respectively.

\paragraph*{Benchmark against experimental $1\times 2$ MMI.} 
In order to benchmark the 2D simulations and the forward neural network performance against an actual device, we consider a simple  1-input by 2-outputs ($1\times 2$) MMI with etched perturbations of index contrast $\Delta n_\text{eff} = -0.71$, corresponding to the earlier described rib waveguides.
We compare formerly designed MMI patterns with the effective index approximation and with the $1\times 2$ forward network.
	
In figure~\ref{fig:experiment_1x2}, we show a comparison of the $1\times 2$ MMI forward network with FDTD simulations and experimental data.
Fig.~\ref{fig:experiment_1x2}a shows a scanning electron microscopy (SEM) image of the patterned MMI device. The device transmission, normalized to a reference straight waveguide, was measured to be $-22.2$\,dB and $-1.0$\,dB (at $\lambda_0=1550$\,nm) for the top and bottom outputs respectively (see supporting information section~G for details and for a comparison with an unperturbed MMI).

Figure~\ref{fig:experiment_1x2}b shows a perturbation-based raster-scan transmittance measurement,\cite{bruckDevicelevelCharacterizationFlow2015} which has been demonstrated to yield a mapping that is proportional to the local electric field intensity integrated over the surface area of the perturbation when the transmittance is nearly unity~\cite{vynckUltrafastPerturbationMaps2018}. This can be seen in the qualitative comparison to the electric field simulation shown in Fig.~\ref{fig:experiment_1x2}c.
	
In figure~\ref{fig:experiment_1x2}d-e we compare the normalized intensity profiles across the output waveguides obtained by experiment (green lines) and FDTD simulation (blue lines) and the forward ANN (orange lines), showing an excellent agreement. Overall this result demonstrates that the 2D simulations as well as the forward network deliver an accurate description of the experimental conditions, which implies the feasibility of an inverse-design ANN to mold the flow of light in integrated photonic devices via optical perturbations, in agreement with earlier studies \cite{bruckAllopticalSpatialLight2016}.

Figure~\ref{fig:experiment_1x2}f shows selected examples of inverse designed patterns from the final iteration of the generator, showing an excellent performance of the ANN-designed MMI patterns in comparison to FDTD simulated intensity-profiles. Target and simulated transmittance values are indicated by black and orange labels in the panels. The average error between design-target and FDTD simulation is generally as low as a few percent. Additional examples of the symmetrical cases are shown in the supporting information figure~S6.

Finally, figure~\ref{fig:experiment_1x2}g compares the peak intensity error between the initial and final iterations of the ANNs for all splitting ratio cases in steps of 5\% and a total target transmittance of 90\%. Note that not only is the performance of inverse designed patterns significantly better after 5 iterations, but also that the agreement between the forward-ANN prediction and simulated response is improved.

\subsection*{Design of transmission matrix in a multi-port MMI}

\begin{figure}[t]
	\centering
	\includegraphics[width=8cm]{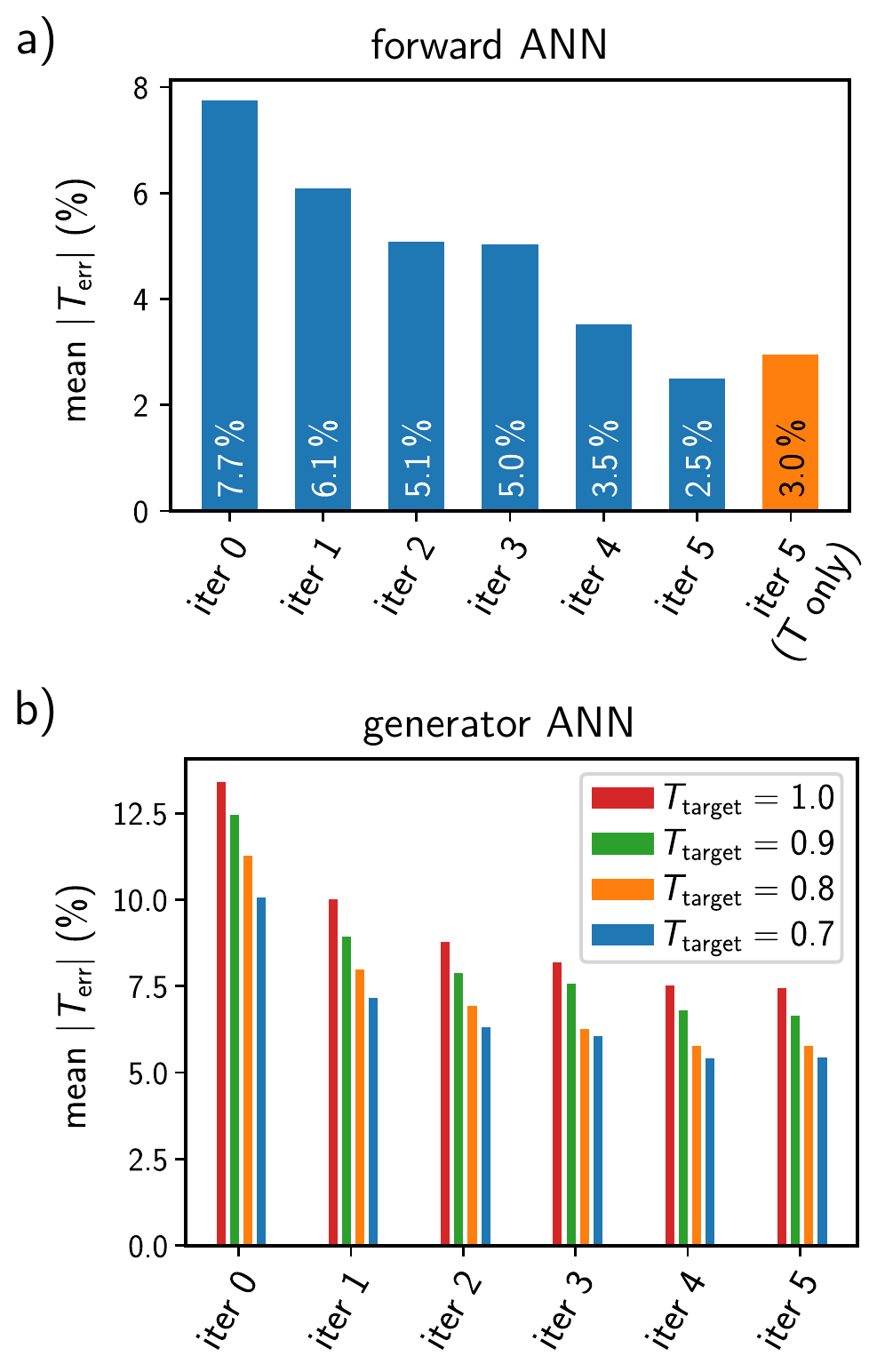}
	\caption{(a,b) 3x3 MMI mean absolute transmittance error between the ANN and simulations for the five consecutive data-generation iterations, calculated on a fixed test dataset.
		(a) forward network error: ANN-prediction vs. simulation.
		Iteration~5 (the full training set) is also compared to an ANN of identical architecture, where the last layer is replaced by a fully connected layer returning only the total transmittance of each port instead of the intensity profile.
		(b) inverse network error: design-target vs. simulation of suggested inverse-design.
		Total transmittance of the design targets is fixed to 70\% (blue), 80\% (orange), 90\% (green) and 100\% (red).
	}
	\label{fig:3x3_inverse_results_mean_Terr}
\end{figure}

\begin{figure*}[t]
	\centering
	\includegraphics[width=0.8\linewidth]{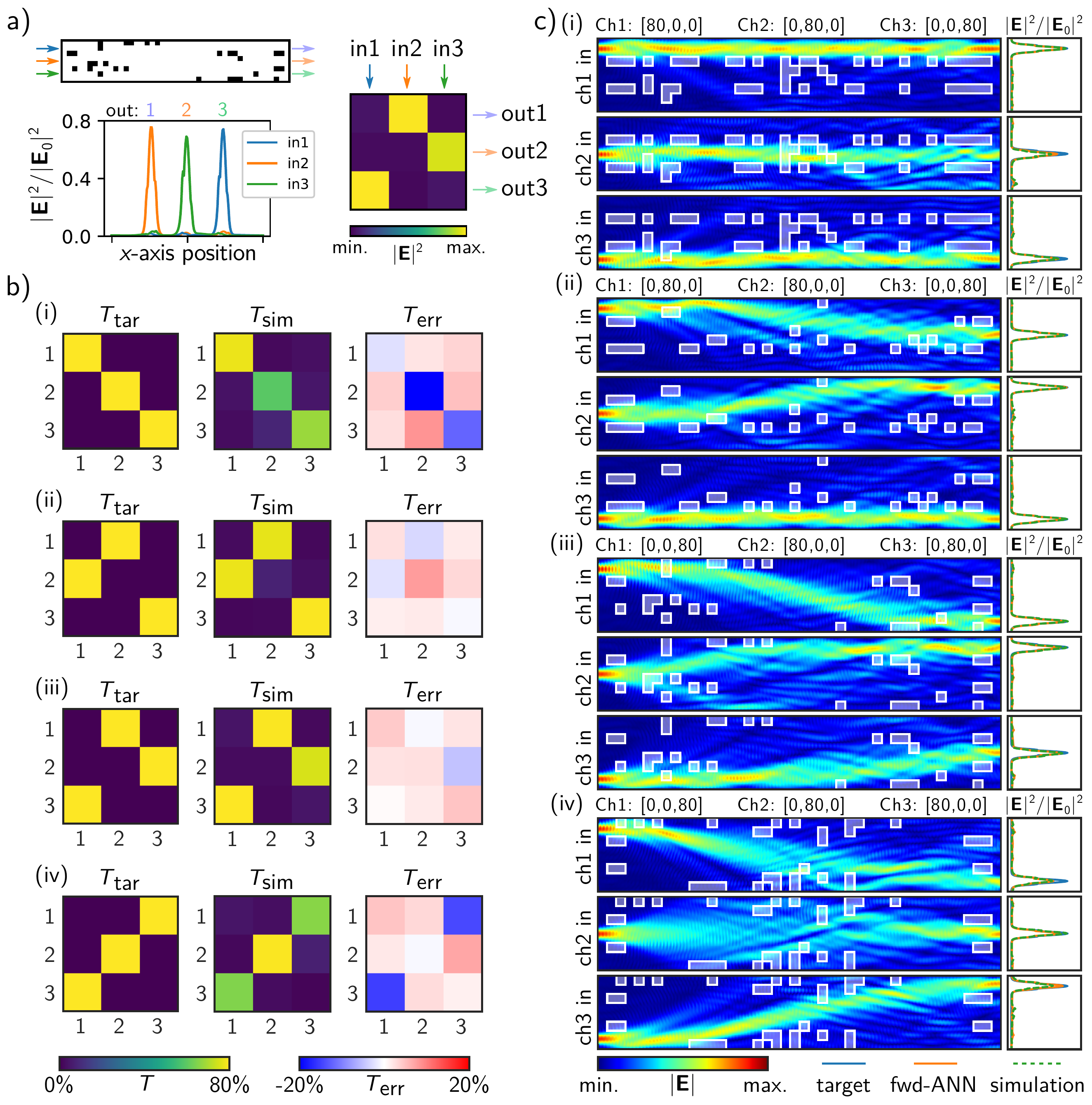}
	\caption{(a) illustration of the used color-array representation.
		The total transmitted intensity for each input/output port couple is represented by a color-code in a small $3\times 3$ color-array.
		Thus, this array represents the intensity transmission matrix of the device.
		(b,c) examples (i-iv) of designed 3$\times$3 transmission matrices (b) and corresponding simulated electric field distributions of the patterned MMI calculated using FDTD (c), for fundamental mode input in the three respective channels. Examples have a target of 80\% total transmittance for each input channel. Intensity transmission matrices are shown for target ($T_{\rm tar}$), simulated design ($T_{\rm sim}$) and error ($T_{\rm err}$). Right side of (c) shows the intensity profiles along the output waveguides for design target (solid blue lines), forward-ANN (solid orange lines) and FDTD simulation (dashed green lines).
	}
	\label{fig:3x3_inverse_results}
\end{figure*}

\begin{figure*}[t]
	\centering
	\includegraphics[width=0.75\linewidth]{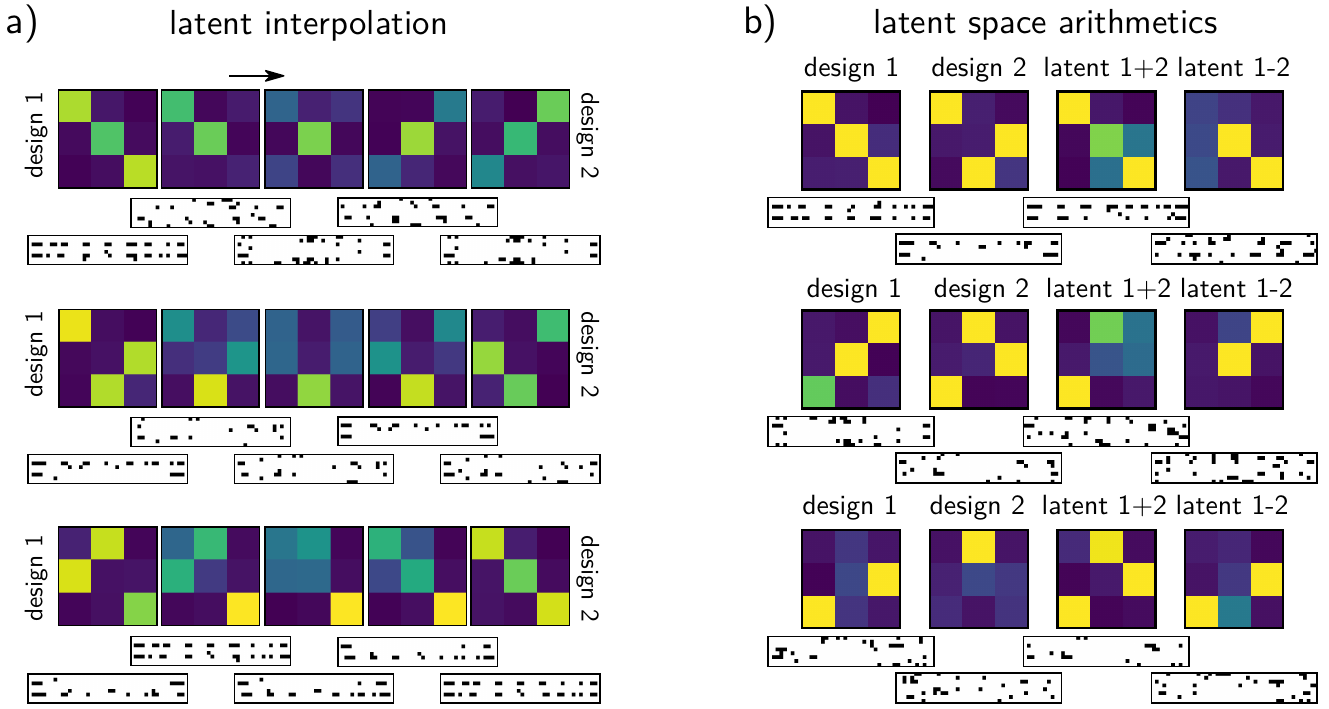}
	\caption{Variational generator latent space operations.
		(a) solutions with a smooth transition of the physical response obtained by linear interpolation of the latent space vector between two inverse-designed MMI patterns (``design 1'' at the left and ``design 2'' at the right end).
		(b) examples of simple arithmetic operations in the latent space of our variational generator ANN.
		This is demonstrated here for three examples of latent-space summation and subtraction.
		All shown color-scale transmission matrices are calculated by FDTD simulations,  color coding is the same as in figure~\ref{fig:3x3_inverse_results}.
	}
	\label{fig:3x3_VAE_latent_operations}
\end{figure*}

\paragraph*{Intensity-only transmission matrix design.} 
The $1\times 2$ MMI discussed so far reflects the simplest possible device configuration and corresponds to the case of a variable power splitter. Having successfully achieved the optimization of this case, we are now in a position to move to the more complex case of multi-port transmission matrices.
In the following we test our inverse ANN on the example of a $3\times 3$ MMI. Instead of using permanent etched perturbations, we consider in the following weaker index-contrast perturbations, $\Delta n_\text{eff} = -0.25$, corresponding to the optically induced perturbations of a recently demonstrated experimental configuration \cite{bruckAllopticalSpatialLight2016}. The effects of individual smaller perturbations is reduced and therefore more perturbations can be added to shape the light flow without strongly affecting the device throughput. Furthermore, we choose the $3\times 3$ MMI to have the same $33$\,\textmu m by $6$\,\textmu m footprint as we use for the $1\times 2$ devices. These dimensions result in very poor unperturbed splitting performance, rendering the design-problem even more challenging, but maintain the compact device footprint. We note that a conventional $3\times 3$ symmetric MMI splitter, i.e. with equal splitting ratios for all port couplings, and the same 6\,\textmu m width requires an 88\,\textmu m device length \cite{Soldano1995Optical}. Such an increased device length renders the transmission more sensitive to small fabrication imperfections and in fact may not be required when using the weak perturbation approach to shape the transmission matrix, provided that the device length is large enough to accommodate the required perturbation patterns.

For this geometry, we first test the performance of both the forward and the inverse networks after each data generation and training cycle. Each such cycle is done on a set of 1000 cases, generated in the same way as during step 4 of the iterative training scheme. For the inverse ANN, we furthermore split our test targets into four different total transmittance ranges with values of 70\%, 80\%, 90\% and 100\%, therefore 250 target transmission matrices are considered for each throughput value. The generated patterns are subsequently simulated by FDTD and results are compared to the target matrices. The accuracy of the forward and inverse ANNs compared to FDTD simulations is shown in figure~\ref{fig:3x3_inverse_results_mean_Terr}a and~b as function of the training iteration number. There is a clear improvement in performance with subsequent iterations.
After the iterative training, both the forward and inverse ANNs have experienced significant performance improvements of a factor~3$\times$ and 2$\times$, respectively, compared to the initial dataset of iteration 0.

In figure~\ref{fig:3x3_inverse_results_mean_Terr}a we additionally compare the intensity-profile predicting forward-network with an ANN that predicts only the ``bare'' $T$ values in each port (orange bar on the very right). We find that training a network on the full spatial intensity distribution instead of the per-port transmittance values leads to a slightly more accurate model ($2.5$\% vs. $3$\%). While this is not a dramatic difference, the inverse training crucially relies on the forward ANN loss, and hence obtaining the best possible accuracy for the forward network is essential for the performance of the inverse network. Therefore, we use the full-profile in this work, but the technique should work also with ANNs operating on transmittance values only.

In figure~\ref{fig:3x3_inverse_results}, we show examples of transmission matrix design. The examples under study are part of an important class of operations, namely those that permutate the device outputs, routing light from each input channel to a separate output port. The routing can then be dynamically controlled via the application of appropriate patterns. Figure~\ref{fig:3x3_inverse_results}a shows how the transmission matrix is constructed from the transmission for each independent input for a particular device. We consider here the intensity transmission matrix between the input and forward-output channels, which is depicted by a color-array, in which each column corresponds to an input channel and each row to an output channel. The possibility of obtaining a full, complex transmission matrix is discussed further below. We do not consider reflections back into the input ports in the matrix, which are generally weak for the regime under study~\cite{bruckAllopticalSpatialLight2016} and which are therefore considered as part of the overall insertion loss of the device. Figure~\ref{fig:3x3_inverse_results}b compares the target and generated matrices $T_{\rm tar}$ and $T_{\rm sim}$, where the error matrix $T_{\rm err}$ gives the differences. We see an overall reduction in the total intensity of the design versus target total transmission, which was set to 80\% (approx. -1\,dB) in this case. Apart from this overall insertion loss, the generator ANN successfully manages to define devices that reproduce the target transmission matrix, as can be seen by the FDTD simulations in figure~\ref{fig:3x3_inverse_results}c.

In the supporting information figure~S7, we show further examples of inverse design for many different input/output routing configurations, demonstrating a very robust and flexible performance of the inverse design ANN.
We note, that the statistics in figure~\ref{fig:3x3_inverse_results_mean_Terr}b are based on randomly generated, arbitrary transmission states, further demonstrating the flexibility of the method.

\begin{figure*}[t]
	\centering
	\includegraphics[width=\linewidth]{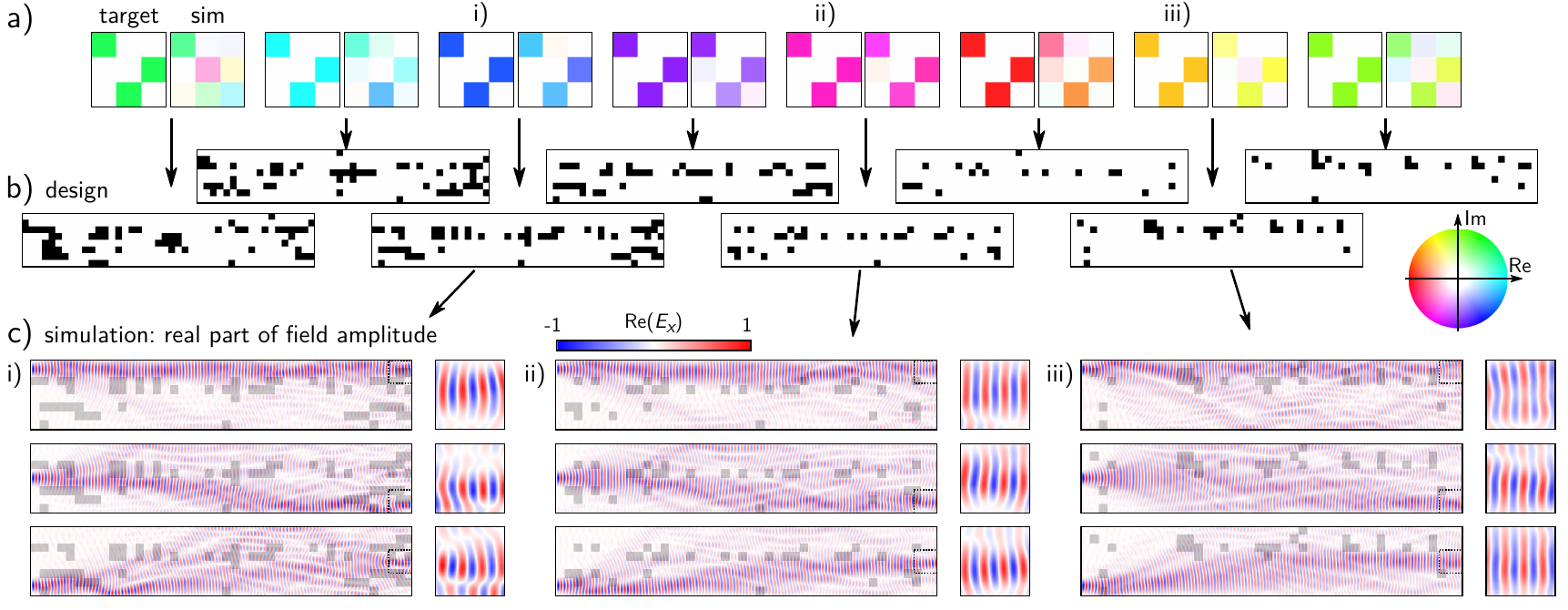}
	\caption{Example of design of a complex transmission matrix.
		An inverse network is trained using internally a full-field forward network and taking as inverse target the real and imaginary part of the complex field amplitude $E_x$.
		This allows to tailor not only the transmitted intensity, but also the relative phase of each transmission channel.
		(a) left: design target. We fix an intensity pattern and rotate the overall relative phase with respect to the input. Right: simulated complex transmission matrix for $E_x$. In order to increase the readability of the figure the colorcode is clipped to white for transmitted intensities $< 5$\%.
		(b) inverse designed perturbation patterns.
		(c) Real part of $E_x$ along the full MMI for the examples i)-iii), where (ii) and (iii) are designed to have respectively a $\pi/2$ and $\pi$ phase relative to (i). The colorscale is normalized to the input amplitude.
		From top to bottom: Input into upper, center and lower channel.
		The right subplots show a zoom of $2\times 2$\,\textmu m$^2$ on the respective target output channel regions (indicated by a dashed square).
	}
	\label{fig:3x3_phase_design}
\end{figure*}

\paragraph*{Latent space operations.} 
During training, the generator network learns a compressed representation of the design targets -- the so-called latent representation. Using the above discussed ``KL-loss'' forces the network to ensure that the latent vectors follow a normal distribution\cite{i.higginsVVAELearningBasic2017}.
In this case, it is possible to perform arithmetic operations such as interpolation, addition or subtraction in the latent space.
Through the interpolation between design targets we can, for instance, obtain MMIs with a smooth transition in their physical behavior, and hence between their transmission matrices.
We illustrate this for several examples in figure~\ref{fig:3x3_VAE_latent_operations}.

In figure~\ref{fig:3x3_VAE_latent_operations}a we interpolate the latent vectors between pairs of generated patterns corresponding to different intensity transmission matrices. The simulated transmission matrices show that the ANN indeed learned a smooth latent representation. Further examples can be found online as movies, in which the transition between two physical solutions is animated along the interpolation.
In figure~\ref{fig:3x3_VAE_latent_operations}b we show three examples of latent space summation and subtraction.
The latent vectors of two designs are added or subtracted and the resulting latent vector is fed into the generator. This yields an MMI with accordingly combined or subtracted transmission matrix.
As can be seen when comparing the MMI patterns of the latent-operations, the solutions do not correspond to a simple summation of, or an interpolation between the MMI patterns.
We note that in case of the latent operations we cannot use the forward network in order to find the best threshold value to get a binary representation of the generated pattern, in the examples of figure~\ref{fig:3x3_VAE_latent_operations} we therefore used the threshold of the respective ``design 1''.

\paragraph*{Phase-aware inverse design.} 
So far, the ANNs in our study were operating only with the transmitted intensity. This is sufficient when considering for example routers where intensity is routed between ports with little cross-interference, or when considering mutually incoherent input channels where coherent cross-interference at the outputs can be neglected. For many applications, such as coherent datacommunication, quantum optics, and analog computing photonic devices\cite{Wang2020, estakhriInversedesignedMetastructuresThat2019}, full control over the complex transmission matrix is required, and hence over both amplitude and phase of the electric field at the outputs. An overall phase offset may be introduced by adding additional phase shifters behind the output ports but this approach is insufficient for the most general case where each output port is coupled to each input with a completely independent phase. Therefore it is of great interest to include phase control into the transmission matrix design itself. To introduce full phase control, we start by training the forward ANN on the prediction of the complex electric fields across the output waveguides obtained using FDTD. The prediction accuracy is found to be similar to the intensity-only forward ANN. However instead of the $M$ ANN-outputs in the intensity-only case, where $M$ is the number of MMI input channels, the complex fields predicting forward ANN returns $4\times M$ profiles, where the factor $4$ reflects the real and imaginary parts of $E_x$ and $E_z$.
Note that in our 2D simulations using TE polarized light, the out-of-plane component $E_y=0$.

For the phase-aware generator model, we assume that $E_x \gg E_z$, in which case the phase of the transverse mode can be approximated by $\vartheta = \text{arctan}(\text{Re}(E_x) / \text{Im}(E_x))$. To avoid having the spatial profiles of complex field amplitudes as design targets, we use for the generator ANN only the real and imaginary parts of the field at the location of peak intensity in the respective output waveguide, $\text{Re}(E_{x, \text{peak }I})$ and $\text{Im}(E_{x, \text{peak }I})$. The ANNs are trained on the design dataset from iteration 5 obtained from the intensity-only $3\times 3$ ANN.

As a proof-of-principle of design of the complex transmission matrix, we select a fixed intensity matrix corresponding to the pattern shown in the left panels of Fig.~\ref{fig:3x3_phase_design}a. Next we define the real and imaginary part of the amplitude such that the output phase is rotated through a $2 \pi$ cycle with respect to the input. The complex amplitude for the MMI transmission is illustrated in Fig.~\ref{fig:3x3_phase_design}a with a hue / brightness based color-coding (see legend on the right). In this example, our target assumes a constant phase for all output channels. In the SI we also show the inverse design of phase-rotation for more complex challenges, where we furthermore add a relative phase of $\pi/6$ between each non-zero output, in order to show independent control of the phase of each port.
This and more examples are given in the supporting information figure~S8.

The MMI patterns obtained by the generator ANN are shown in Fig.~\ref{fig:3x3_phase_design}b, the corresponding simulated complex amplitude transmission matrices are depicted in Fig.~\ref{fig:3x3_phase_design}a in the right panels.
To improve the readability of the figure, we clip the color-coding to white for output-channels with transmitted intensities $<5\%$.

We observe that in general, the inverse network succeeds in finding MMI patterns which reproduce both the desired transmission pattern as well as the target phase at the outputs.
Examining the examples labeled (i) to (iii) in Fig.~\ref{fig:3x3_phase_design}c, we observe that the ANN uses two basic strategies to achieve phase design. The first strategy consists of the introduction of perturbations into or close to the light path, which induces a phase shift due to the effective index change of the propagation medium. The second strategy involves the modification of the light path length.
In the latter case, the ANN gains control over the phase by routing the light along shorter or longer paths, as can be observed in the center and bottom input cases in Fig.~\ref{fig:3x3_phase_design}c.

In most cases, the phase-design ANN yields MMI patterns with reasonable qualitative performance. However we note that the agreement is quantitatively much worse than the above shown intensity-only examples. Controlling both phase and intensity is a far more complex problem than intensity only transmission design.
Hence, full fine-control over the phase would probably require a larger MMI footprint and/or a larger number of stronger perturbations. While there are no principle technical constraints, larger MMIs and smaller perturbations would considerably increase FDTD simulation time and hence slow down data generation and the training phase considerably. Further studies are needed to explore this design space.

Despite the generally good qualitative performance, in certain cases the inverse-designed MMIs perform poorly where some channels suffer from low transmission. This can be seen for instance in the very left design target of figure~\ref{fig:3x3_phase_design}, where the number of perturbations in the ANN design is very large. But also in these cases of poor inverse design performance, we observe that the forward network predictions are in very good agreement with the FDTD simulations. Therefore we conclude that the phase-design problems are not due to a poor generalization capability of the ANNs. Judging from the perturbation patterns (Fig.~\ref{fig:3x3_phase_design}b and also several examples in supporting information Fig.~S8), we believe that the problematic cases are at the very limit of what is possible within the relatively small MMI-size and with the limited amount and weak index contrast of the perturbations. In case of the very left MMI in figure~\ref{fig:3x3_phase_design}, the ANN adds a large amount of perturbations in an attempt to tune the phase to the desired angle. The high perturbation density however deteriorates severely the general performance of the device, suppressing the transmittance for input channels 2 and 3.
In the MMI at the very right of figure~\ref{fig:3x3_phase_design}, the ANN followed the opposite strategy and inserted as few perturbations as possible, obviously arriving at the other side of the limit of its working range.
In agreement with those observations we find that the phase-design MMI patterns with the largest and with the least number of perturbations usually perform worst (c.f. supporting information Fig.~S8).
The deterioration of the MMI performance with too many or too few perturbations can also be seen in the supporting information figure~S3.

\section*{Discussion}

Our investigations have shown that neural networks can be trained to generate arbitrary transmission matrices by designing a corresponding perturbation pattern in a multi-mode device. One of the main challenges in this work was to give the network sufficient examples of high-throughput solutions, which are the desirable patterns for many applications but are also sparsely distributed in the overwhelmingly large design space.
	
A valid question is whether the neural network actually conceives a generalized representation or if it merely `memorizes' results through an elaborate lookup table functionality. To determine whether the neural networks generalize well and find original solutions beyond what is available in the simulated database, we compare the inverse designs of our test-cases with the training data.
In table~S1 presented in the supporting information we summarize the average number of non-identical perturbations between the ANN designs and the respective most similar patterns in the training data, both for the ANNs trained on only the initial data set as well as for the ANN after the iterative training.
In case of the $1\times 2$ MMIs we observe that after the iterative training, for any inverse designed pattern there can be found in most cases an almost identical device in the training set ($<2$ non-matching perturbation positions).
We attribute this overlap to the lower complexity of the $1\times 2$ MMI problem, being essentially just a power-splitter.

For the more complex $3\times 3$ MMIs on the other hand, we found that the ANN does not simply act as a lookup table and the generated designs are usually very different from the closest design in the dataset (average mismatch count $17.1 \pm 12.2$ after iterative training). In this case, the number of non-matching perturbations is not decreasing significantly during the iterative training. Clearly, in order to solve the rather complex $3\times 3$ transmission matrix problem, it is necessary for the ANN to generalize well, in order to be able to invent original solutions.

For the phase-aware transmission matrix design, we have shown the potential as well as the limitations for the current geometry. Overall the neural network is capable of including phase in the transmission matrix. In the supporting information figure~S8 we show additional examples for the phase-design network, including an example with the same intensity-target as in figure~\ref{fig:3x3_phase_design}, but with independent phases for the different output channels as compared to the same overall phase. A similar performance is achieved, indicating that there is no constraint in the design of the output phases for each port.

Again, as in the above discussed examples, the cases of poor MMI performance can usually be explained by insufficient design margins in our chosen geometry which limits the ANN in tuning the phase over a full $2\pi$ range. Larger device geometries and/or stronger perturbations are expected to provide access to a larger design space, however, this needs to be traded off against transmission losses which increase with the number of perturbations and their strength. While current work has been aiming to achieve high-throughput solutions, for real-world applications the extinction ratio, crosstalk between ports, or robustness against structural imperfections will be additional important parameters, which cannot be easily implemented as optimization targets in the neural network approach. We anticipate that these additional performance figures can be fine tuned, for example by further improving the solutions generated by the network using multi-objective iterative forward optimization.\cite{piggottFabricationconstrainedNanophotonicInverse2017}

In general, an important disadvantage of deep neural network based, direct inverse design is the high computational cost, associated with training data generation. Therefore, in applications where only a few device designs are required, it might be preferable to use conventional methods such as adjoint optimization or evolutionary algorithms.\cite{piggott2015inverse, wiechaEvolutionaryMultiobjectiveOptimization2017} 
The conventional methods furthermore do not introduce statistical errors as ANN-inverse design does. The latter is data-driven and therefore implies inter- and extrapolation errors which are usually in the order of a few percent.
On the other hand, once trained the ANN delivers the inverse design in milliseconds, whereas conventional approaches usually require hundreds to tens of thousands of simulation evaluations per design, which reflects in hours or even days of runtime.
Furthermore, if several objectives are simultaneously target of a design problem, conventional optimization methods can be difficult to employ, requiring the careful definition of a multi-objective fitness function, or a non-trivial Pareto optimization.
Therefore, an important advantage of our deep learning approach is its capability to directly solve multi-objective problems such as the above demonstrated concurrent design of 3 distinct input situations with 3 output channels, including both phase and amplitude targets.

It is also important to mention that our approach, such as most deep learning techniques, requires a completely new dataset for each new design geometry, i.e. after variations in length or width of the MMI, or of the perturbation strength. 
This large computational overhead limits the range of parameters that can be easily explored, reducing the flexibility of the method. However, the required data to train networks on new, yet similar problems might be significantly reduced using procedures such as transfer learning, whereas variable input dimensions might be realized with so-called \textit{attention}-based concepts.\cite{quMigratingKnowledgePhysical2019, luoProbabilityDensityBasedDeepLearning2020} 
According research in the context of photonics is still scarce, hence such techniques are interesting as subject of follow-up work.
Future work will also look at expanding the geometrical design space, which includes increasing the number of input and output channels to ultimately achieve large scale multi-input, multi-output (MIMO) programmable devices. Such explorations may be able to achieve further improvements in the neural network approach, including aforementioned strategies such as transfer learning to improve the agility of the neural approach in adapting to variations in device geometries.

\section*{Conclusion}

In conclusion, we have developed a method of designing universal optical components based on weak scattering perturbations in a multimode waveguide. Arbitrary transmission matrices were generated using a deep learning neural network. By directing the neural network toward high throughput solutions using an iterative data generation scheme, we have achieved reliable results with a mean-square error $<$5\% compared to numerical simulations. The neural network approach, while computationally demanding to train, is fast ($<$50~ms) in the generation of solutions and hence could be used in real-time applications. We have furthermore shown that independent phase control can be readily implemented, where we have seen some of the limitations in the limited design space under study. This work opens up a conceptually new approach for programmable photonic systems based on weak and forward multiple scattering effects. Such integrated devices with a freely tunable transmission matrix, at length scales far below that of interferometer meshes, will find application as ultracompact universal optical elements for applications in optical routing, programmable weight banks, or analogue computing.

\section*{Methods}

\subsection*{ANN details and hyperparameters}

We use a tandem network consisting of two networks, as shown in figure~\ref{fig:training_scheme}b in the main text.
The first part is a forward network (green box), which is pre-trained in a first step to predict the MMI transmission. 
The second constituent is the variational generator network (blue box), which shall generate MMI patterns that fulfil the desired transmittance target.
Both networks are encoder-decoder ResNet models\cite{szegedyInceptionv4InceptionResNetImpact2016}, consisting of down- and up-sampling residual blocks. Each residual block consists of three repetitions of consecutive convolutional, batch normalization and leaky ReLU layers. After each residual block, a max pooling operation is performed to divide the layer size by two. After each max pooling the number of filters is doubled, following the commonly used design principle.
The output layer of both the forward network and of the generator uses a sigmoid activation to match the numerical range of the normalized intensities or field amplitudes, respectively the binary MMI patterns (zero or one).
In the decoders, zero padding is used when necessary to reconstruct a network output of correct dimensions. This is indicated as ``(+1)'' in the labels indicating layer dimensions in figure~\ref{fig:training_scheme}b of the main text.
The kernel sizes are in the 1D case for decreasing layer dimension: 5, 4, 3 and 3. 
In the 2D parts of the networks the kernel sizes are $(4\times 4)$, $(3\times 3)$, $(3\times 3)$ and $(2\times 2)$. 
In the forward network, a dense layer of $256$ neurons interconnects the 2D encoder and the 1D decoder.
In the case of the variational generator we use a latent vector representation of length $64$. We use the common reparametrization trick\cite{kingmaIntroductionVariationalAutoencoders2019} in order to be able to backpropagate through the latent representation despite the random sampling of latent vectors $z$, which is required to calculate the KL-loss. 
To this end, the network is built such that it represents a latent vector using its mean $\mu$ and variance $\sigma$, which are differentiable for backpropagation. From those values we can then draw random samples assuming a normal distribution $N(\mu, \sigma)$.

We first train only the forward network (step 2 of Fig.~\ref{fig:training_scheme}a), for which the ANN-predicted intensity-profiles are compared to the numerically simulated MMI output. Training is done using the ADAM solver\cite{kingmaAdamMethodStochastic2014} with a learning rate of $5\times 10^{-5}$ and a batchsize (BS) of 32. 
We increase the BS by a factor of 2 after 25 epochs and by another factor of 2 after 60 epochs.\cite{smithDonDecayLearning2018}
Once the forward network was trained for 100 epochs, we fix its parameters and train the full model, using a combination of three loss functions: first the MMI pattern loss (comparing generated MMI and training sample MMI patterns with a mean-square error (MSE) metric), second the $|\mathbf{E}|^2$-profile loss (comparing the pre-trained forward network's prediction of the generated MMI with the training sample's intensity profile on an MSE metric), and third the KL-loss to assure a uniform latent space\cite{kingmaIntroductionVariationalAutoencoders2019}.
The three losses are weighted by $1:50:0.2$, which we determined to work well. 
Furthermore, in order to assure stable training, we found it to be necessary to ramp up the weight of the KL loss from zero to its final value $0.2$ over the first 30 epochs (``KL-annealing''\cite{bowmanGeneratingSentencesContinuous2016}).
We use the same learning rate and BS increase scheme for the forward network alone as well as for the full model.

When using the ANN to generate new patterns, the total transmittance target has a direct effect on the networks performance. Therefore, each input channel is assigned a randomized total transmittance between 70--100\%, which is then randomly distributed over the different output ports.
To satisfy reciprocity the transmission matrices need furthermore to be unitary, which we obtain from the target matrix elements by also re-normalizing the row vectors (output-channels) in case their norm is $>1$.

\subsection*{Sample fabrication}

Devices were fabricated from a silicon-on-insulator (SOI) wafer (220\,nm thick silicon layer, 2 micron thick buried oxide) using e-beam lithography on a 265\,nm thick layer of ZEP photoresist and subsequent alcatel inductive coupled plasma-reactive ion etching of the silicon.
A 120\,nm etch depth was used to form the single mode 500\,nm wide rib waveguides which are tapered up to 1\,micron at the MMI boundaries, ensuring an adiabatic size conversion of the fundamental TE mode over a 10\,micron distance. Grating couplers were used to allow the coupling of light into and out of the device.

\subsection*{Ultrafast perturbation-based transmittance measurement}

A mode-locked Ti:Sapphire laser operating at 830\,nm was used in conjunction with an optical parametric oscillator to provide 200\,fs probe pulses centered around 1550\,nm at a 80\,MHz repetition rate. 
The probe light was coupled into and out of the device using optical fibers and grating couplers, and the transmission was recorded by an InGaAs avalanche photodiode connected to a lock-in amplifier. The second harmonic of the seed laser was used as the optical pump at 415\,nm wavelength and its repetition rate was halved to 40\,MHz using a pulse picker. 
The pump was focused onto the surface of the device using a 50x objective with a numerical aperture of 0.55, resulting in a perturbation spot size of 740\,nm (full-width half-maximum) at the focus. 
Optical fluence was set to 20\,pJ/\textmu m$^2$ which results in an effective index shift of $\Delta n_{\text{eff}} = -0.1$ due to free-carrier excitation in the silicon.
A variable time delay stage enabled the precise temporal overlap of the two pulses, which was set to have a 3\,ps probe delay, and a three-axis nanopositioner stage was used to raster scan the pump over the device. The differential transmission value, $\Delta T/T$, at each position was obtained by using the lock-in amplifier to demodulate the transmission signal at both pulses' repetition frequencies.

\vspace{2\baselineskip}

\noindent \textbf{Supporting Information} This material is available free of charge via the internet at \href{http://pubs.acs.org}{http://pubs.acs.org}. 
Additional details on data generation, network layout and training procedure.
Explanations on the binarization of network generated MMI-patterns.
Detailed statistics of the ANN performance for the different MMI models and comparison between inverse designs and training samples.
Further inverse design examples.

\noindent \textbf{Acknowledgements:}
We thank the NVIDIA Corporation for the donation of a Quadro P6000 GPU used for this research.\\

\noindent \textbf{Funding:} Silicon photonic waveguides were manufactured through the UK Cornerstone open access Silicon Photonics rapid prototyping foundry through the EPSRC grant EP/L021129/1.
P.R.W. acknowledges support by the German Research Foundation (DFG) through a research fellowship (WI 5261/1-1) and the computing facility CALMIP under grant p20010.
OM acknowledges support through EPSRC grant EP/M009122/1.\\

\noindent \textbf{Competing Interests} The authors declare that they have no competing financial interests.\\

\noindent \textbf{Data and materials availability:} All data supporting this study are openly available from the University of Southampton repository (DOI: 10.5258/SOTON/XXXXXXXXX).

\bibliographystyle{achemso}
\bibliography{DinsdaleMMI2020.bbl}

\end{document}